\documentclass[onecolumn,aps,pra,showpacs,floatfix,superscriptaddress]{revtex4}

\usepackage{amsmath}
\usepackage{amsfonts}
\usepackage{amssymb}
\usepackage{bm}
\usepackage{bbm}
\usepackage{graphicx}
\usepackage{color}
\usepackage{maybemath}
\usepackage{fancybox}

\newcommand{\mat}{\bm}
\newcommand{\vvv}{\bm}

\newcommand{\dd}{\mathrm{d}}
\newcommand{\ee}{\mathrm{e}}
\newcommand{\ii}{\mathrm{i}}

\newcommand{\half}{\tfrac12}

\renewcommand{\Re}{\mathrm{Re}}

\newcommand{\bT}{{\mat T}}
\newcommand{\bG}{{\mat G}}
\newcommand{\bL}{{\mat L}}

\newcommand{\bZ}{{\mat Z}}
\newcommand{\bH}{{\mat H}}
\newcommand{\bJ}{{\mat J}}
\newcommand{\bA}{{\mat A}}
\newcommand{\bB}{{\mat B}}

\newcommand{\bQ}{{\mat Q}}
\newcommand{\bx}{{\vvv x}}
\newcommand{\by}{{\vvv y}}
\newcommand{\bv}{{\vvv v}}
\newcommand{\bu}{{\vvv u}}

\newcommand{\be}{\hat{\mat  e}}

\newcommand{\rmT}{{\rm T}}

\newcommand{\scp}[2]{\langle #1,#2 \rangle}

\newcommand{\calP}{\mathcal{P}}

\newcommand{\calT}{\mathcal{T}}

\newcommand{\PT}{$\mathcal{PT}$}

\newcommand{\plus}{{\mbox{{\bf{\tiny +}}}}}

\begin{document}

\title{Generalized Householder Transformations\\
for the Complex Symmetric Eigenvalue Problem}

\author{J. H. Noble}
\affiliation{Department of Physics,
Missouri University of Science and Technology,
Rolla, Missouri 65409, USA}

\author{M. Lubasch}
\affiliation{Max--Planck--Institut f\"{u}r Quantenoptik,
Hans--Kopfermann--Stra\ss{}e 1, 85748 Garching, Germany}

\author{U. D. Jentschura}
\affiliation{Department of Physics,
Missouri University of Science and Technology,
Rolla, Missouri 65409, USA}
\affiliation{MTA--DE Particle Physics Research Group, 
P.O.Box 51, H-4001 Debrecen, Hungary}

\begin{abstract}
We present an intuitive and scalable algorithm
for the diagonalization of complex symmetric matrices,
which arise from the projection of 
pseudo--Hermitian and complex scaled Hamiltonians
onto a suitable basis set of ``trial'' states.
The algorithm diagonalizes complex and symmetric
(non--Hermitian) matrices and is easily implemented
in modern computer languages.
It is based on generalized Householder 
transformations and relies on iterative similarity transformations
$\bm{T} \to \bm{T}' = \bm{Q}^{\rm T} \, \bm{T} \, \bm{Q}$,
where $\bm{Q}$ is a complex and orthogonal, but not unitary, matrix, i.e,
$\bm{Q}^{\rm T} = \bm{Q}^{-1}$ but $\bm{Q}^\plus \neq \bm{Q}^{-1}$.
We present numerical reference data to support the 
scalability of the algorithm.
We construct the generalized Householder transformations
from the notion that the conserved scalar product of eigenstates
$\Psi_n$ and $\Psi_m$ of a pseudo--Hermitian quantum mechanical
Hamiltonian can be reformulated in terms of the 
generalized indefinite inner product
$\int \dd x \, \Psi_n(x,t) \, \Psi_m(x,t)$,
where the integrand is locally defined,
and complex conjugation is avoided.
A few example calculations are described which illustrate 
the physical origin of the ideas used in the 
construction of the algorithm.
\end{abstract}


\maketitle

%
%
\section{Introduction}

\indent Complex symmetric matrices $\bm{A} = \bm{A}^{\rm T}$ arise naturally from the
projection of a complex scaled (```resonance-generating'') Hamiltonian onto a
basis of quantum mechanical ``trial'' states.  For suitably chosen parameters,
the diagonalization of a matrix of this type leads to (accurate) approximations
for the resonance energies, and the resonance eigenstates, of the complex
scaled Hamiltonian~\cite{JeSuLuZJ2008,JeSuZJ2009prl,JeSuZJ2010,ZJJe2010jpa}.
The resonance energies are manifestly complex; the width of the quantum state
enters the complex energy as $E = \Re \, E - \frac{\ii}{2} \Gamma$, where
$\Gamma$ is the width (we use natural units with $\hbar=c=\epsilon_0=1$
throughout this article).  A paradigmatic example~\cite{JeSuLuZJ2008} is the
complex-scaled cubic anharmonic oscillator Hamiltonian $h_3 = -\frac12 \,
\exp(-2\ii\theta) \, \partial_x^2 + \frac12 \, \exp(2\ii\theta) \, x^2 + G \,
\exp(3\ii\theta) \, x^3$, where $G$ is a coupling parameter and $0 < \theta <
\pi/5$ is a complex rotation angle.

However, complex scaled Hamiltonians are not the only source of complex
symmetric matrices in theoretical physics.  For example, if one projects the
pseudo--Hermitian (\PT-symmetric) anharmonic
oscillator~\cite{BeBo1998,BeBoMe1999} Hamiltonian $H_3 = - \frac12 \,
\partial_x^2 + \frac12 \, x^2 + \ii \, G \, x^3$ with an imaginary cubic
perturbation ($G > 0$) onto a basis of harmonic oscillator eigenstates, then
one obtains a complex symmetric (but not Hermitian) matrix, the eigenvalues of
which are real.

As shown in Refs.~\cite{JeSuZJ2009prl,BeDu1999}, the complex resonance energies
of the ``real'' cubic anharmonic oscillator $h_3$ are connected with the real
eigenenergies of the imaginary cubic perturbation $H_3$ via a dispersion
relation.  The same holds true for all anharmonic oscillators of odd degree.
Some of the numerical calculations were instrumental in providing additional
evidence for the generalization~\cite{JeSuZJ2009prl} of the so-called
Bender--Wu formulas~\cite{BeWu1969,BeWu1971}, which describe the large-order
asymptotic growth of the perturbative coefficients of an arbitrary energy
eigenvalue of an even anharmonic oscillator, to odd anharmonic oscillators.
Indeed, the conjectures on nonperturbative quantization conditions, described
in Refs.~\cite{JeSuZJ2009prl,JeSuZJ2010} had been checked against
high-precision numerical data before the results were presented.

The purpose of this paper is threefold: First, to illustrate the numerical
procedures underlying the numerical verification of the conjectured generalized
quantization conditions, second, to describe an intuitive and scalable (in
terms of the numerical precision) matrix diagonalization algorithm which seems
to be particularly suited for the treatment of complex symmetric matrices.  Our
algorithm has a certain ``twist'' in the sense that it is based on generalized
Householder transformations.  The generalized Householder matrices $\bH_\bv$
have manifestly complex entries but are not Hermitian unlike the familiar
formalism (see p.~225 of Ref.~\cite{StBu2002}).  Instead, they are orthogonal
matrices with the property $\bH_\bv \, \bH^{\rm T}_\bv = \mathbbm{1}$.  To the
best of our knowledge, these generalized Householder reflections have not
appeared in the standard
literature~\cite{Wi1965,WiRe1971,GovL1996,Pa1998matrix,AnEtAl1999} on matrix
diagonalization procedures before.

Finally, the third purpose of the paper is to illustrate a few properties of
the eigenenergies of pseudo--Hermitian Hamiltonians, based on calculations done
with the algorithm presented here. Let us anticipate one of the observations
made in the course of the calculations.  At face value, the conserved scalar
product~\cite{BeBrReRe2004} (under the time evolution governed by a
pseudo--Hermitian anharmonic oscillator) is given as the integral $\int \dd x
\, \Psi_n^*(-x,t) \, \Psi_m(x,t)$; this expression involves a non-local
integrand with function evaluations at $x$ and $-x$.  Typically, eigenstates of
complex symmetric matrices are orthogonal with regard to a conceptually much
simpler scalar product, namely, the indefinite inner product $\int \dd x \,
\Psi_n(x,t) \, \Psi_m(x,t)$, where the integrand is locally defined, and
complex conjugation is avoided. However, the two scalar products are related,
as described in this article, and this observation has motivated the
construction of the matrix diagonalization algorithm presented here.  We thus
proceed by describing the algorithm in Sec.~\ref{sec2} and the physical
motivation for its development in Sec.~\ref{sec3}, together with a few example
calculations.  Conclusions are reserved for Sec.~\ref{sec4}.

%
%
\section{Complex Symmetric Eigenvalues and Eigenvectors}
\label{sec2}

%
%
\subsection{Orientation}
\label{sec2_1}

Quantum mechanics is formulated in an infinite-dimensional vector space
(Hilbert space) of functions.  However, once a numerical evaluation of
eigenvalues of a particular Hamiltonian is pursued, the quantum mechanical
Hamiltonian needs to be projected onto a suitable basis set of wave functions,
leading to a finite-dimensional matrix.

When a pseudo-Hermitian Hamiltonian such as the imaginary cubic oscillator is
projected onto a basis set consisting of harmonic-oscillator eigenfunction, one
obtains a complex symmetric (not Hermitian!) matrix.  Typically, the
generalized indefinite inner product~\cite{GoLaRo1983,Mo1998,ArHo2004} 
naturally emerges as a tool in the analysis of
pseudo--Hermitian quantum mechanics, and one would thus naturally assume that
the indefinite inner product might be useful in the development of a suitable
matrix diagonalization algorithm.  Indeed, the approximate calculation of
eigenvalues of quantum mechanical Hamiltonians naturally emerges as a task in
the analysis of the quantum dynamics induced by the pseudo--Hermitian time
evolution. 

Such an algorithm will be described in the 
following; it essentially relies on two steps:
In the first, the complex symmetric input matrix is 
transformed to tridiagonal form, and in the second step, 
the tridiagonal matrix is diagonalized to machine accuracy.
The first step uses the concept of the complex
inner product in an absolutely essential manner;
it is based on generalized Householder reflection matrices.
The complex symmetric input matrix is transformed to tridiagonal
form in a single computation whose
computational cost is of order $n^2$
(here, ``tridiagonal form'' refers to a form where the 
diagonal, as well as the sub- and superdiagonal entries of the 
matrix are nonzero).
For the second step, one has a number of methods 
available; we shall briefly outline a method based 
on an iterative QL decompositions with an implicit 
(Wilkinson) shift.

%
%
\subsection{Tridiagonalization}
\label{sec2_2}

We first define the indefinite inner product for
finite-dimensional $n$-vectors as follows,
\begin{equation}
\label{gen_inner}
\scp{\bx}{\by}_* = \sum_{i=1}^n x_i \, y_i = 
\bx^{\rmT} \cdot \by \,,
\end{equation}
where $\bx^{\rmT}$ is a row vector, whereas
$\by$ is a column vector.
Note that the entries of $\bx$ and $\by$ may be complex numbers,
but complex conjugation of either $\bx$ or $\by$  is avoided.
We use generalized Householder reflection matrices $\bH_\bv$,
which have the properties,
\begin{align}
\label{gen_house}
\bH_\bv =& \; \mathbbm{1} -
\frac{2}{\scp\bv\bv_*} \, \bv \otimes \bv^{\rmT} \,,
\qquad
\bH_\bv \; \bx= \bx-2 \bu \left< \bu, \bx \right>_* \,,
\qquad
\bu = \frac{\bv}{|\bv|_*} \,, 
\qquad
|\bv|_* = \sqrt{\scp\bv\bv_*} \,.
\end{align}
Here, the dyadic product of a column and a row vector 
is denoted by the symbol $\otimes$, and the branch cut of the square root function 
in the calculation of $|\bv|_*$ is along the negative real axis.

The generalized Householder reflection matrix 
$\bH_{\bv}$ is symmetric, i.e., $\bH_\bv=\bH_\bv^\rmT$.
Furthermore, the Householder reflections are 
square roots of the unit matrix,
\begin{align}
\bH_\bv^2=& \; \bH_\bv \; \bH_\bv^\rmT = \bH_\bv^\rmT \; \bH_\bv
= \mathbbm{1}-4\bu \otimes \bu^T +
4\,\bu \otimes \bu^T \; \left< \bu, \bu \right>_* =\mathbbm{1}.
\end{align}
A characteristic property of the Householder reflections is that the 
parameter vector $\bv$ can be adjusted so that the input vector $\bx$ 
is projected onto a particular axis, upon the calculation 
of $\bH_\bv \; \by$. We set 
\begin{equation}
\bv=\by+|\by|_* \; {\be}_n \,,
\end{equation}
where $\be_n$ is the ``last'' unit vector in the 
$n$-dimensional space, and we verify that 
\begin{align}
\bH_\bv \; \by =& \;
\by-\frac{2}{\left< \bv, \bv \right>_*}\,
\left< \bv, \by \right>_* \, \bv
= \by - \frac{2\left(\by^\rmT+|\by|_* \; \be^\rmT_n\right) \cdot \by}
{\left(\by^\rmT+|\by|_* \, \be^T_n\right) 
\cdot \left(\by + |\by|_*\be_n\right)}\bv
\nonumber\\
=& \; \by -\frac{2\left(|\by|_*^2+|\by|_*y_n\right)}
{2|\by|_*^2+2|\by|_*y_n}\bv
=\by-\bv = -|\by|_*\be_n \,.
\label{propHR}
\end{align}
Furthermore,
\begin{equation}
\by^\rmT\bH_\bv = (\bH_\bv\by)^\rmT = -|\by|_*\be_n^\rmT\;.
\label{propHRT}
\end{equation}
The results from Eqs.~\eqref{propHR} and~\eqref{propHRT}
are useful in the tridiagonalization procedure.
Let $\bA$ be the matrix we want to tridiagonalize.
In the first step, we choose the column vector $\by_{n-1}$ to 
consist of the first $n-1$
elements of the last column of $\bA$,
\begin{equation}
\by_{n-1}=
\left(
\begin{array}{c}
A_{1\,n}\\
A_{2\,n}\\
\vdots\\
A_{n-1\,n}
\end{array}
\right).
\end{equation}
By defining $\bB_{n-1}$ as an $(n-1)\times (n-1)$ matrix
where $B_{ij}=A_{ij}$, for $i,j=1,\dots, n-1$, we can write $\bA$ as
\begin{equation}
\bA=
\left(
\begin{array}{ccc|c}
&&&\\
&\bB_{n-1}&&\by_{n-1}\\
&&&\\
\hline
&
\rule[-3mm]{0mm}{8mm}
\by_{n-1}^\rmT&&A_{nn}
\end{array}
\right) \, .
\end{equation}
In the spirit of the Householder reflection, we set
\begin{equation}
\bv_{n-1} = \by_{n-1}+|\by_{n-1}|_* \, \be_{n-1}\,.
\end{equation}
We can then construct $\bH_{\bv_{n-1}}$, which will be a Householder matrix
of rank $n-1$. The complex $n \times n$ matrix $\bH_{n-1}$ 
is defined as 
\begin{equation}
\bH_{n-1}=
\left(
\begin{array}{ccc|c}
&&&\\
&\bH_{\bv_{n-1}} && \vec 0 \\
&&&\\
\hline
&
\rule[-3mm]{0mm}{8mm}
\vec 0^\rmT&& 1 
\end{array}
\right) \, .
\end{equation}
Then,
\begin{align}
\bA'&=\bH_{n-1}\;\bA\;\bH_{n-1}
= \left(
\begin{array}{ccc|c}
&&&\\
&\bH_{\bv_{n-1}}\bB_{n-1}\bH_{\bv_{n-1}}&&\bH_{\bv_{n-1}}\by_{n-1}\\
&&&\\
\hline
&\by_{n-1}^\rmT\bH_{\bv_{n-1}}&&A_{nn}
\end{array}
\right).
\end{align}
Using Eq.~\eqref{propHR} and~\eqref{propHRT}, this can be reduced to
\begin{equation}
\bA'=
\left(
\begin{array}{ccc|c}
&&&0\\
&\bB'_{n-1}&&\vdots\\
&&& -|\by_{n-1}|_*\\
\hline
0&\cdots& -|\by_{n-1}|_* & A_{nn}
\end{array}
\right),
\end{equation}
where
\begin{equation}
\bB'_{n-1} = \bH_{\bv_{n-1}}\bB_{n-1}\bH_{\bv_{n-1}}\,.
\end{equation}
For the second step we choose $\by_{n-2}$ to be the first $n-2$
elements of the second to last column of $\bA'$,
\begin{equation}
\by_{n-2}=
\left(
\begin{array}{c}
A'_{1\,n-1}\\
A'_{2\,n-1}\\
\vdots\\
A'_{n-2\,n-1}
\end{array}
\right) \,,
\quad
\bv_{n-2} = \by_{n-2}+|\by_{n-2}|_* \, \be_{n-2} \,.
\end{equation}
The Householder matrix $\bH_{\bv_{n-2}}$ 
is of rank $n-2$, and $\bH_{n-2}$ is defined as 
\begin{equation}
\bH_{n-2}=
\left(
\begin{array}{ccc|c}
&&&\\
&\bH_{\bv_{n-2}} && \vec 0 \\
&&&\\
\hline
&
\rule[-3mm]{0mm}{8mm}
\vec 0^\rmT&& \mathbbm{1}_{2\times2} 
\end{array}
\right) \, .
\end{equation}
We write $\bA'$ as follows,
\begin{equation}
\bA' =
\left(
\begin{array}{ccc|c|c}
&&&&\\
&\bB_{n-2}&&\vvv y_{n-2} & \vvv 0\\
&&&&\\
\hline
\rule[-3mm]{0mm}{8mm}
& \vvv y^\rmT_{n-2} & & A'_{n-1\,n-1}& -|\by_{n-1}|_* \\
\hline
\rule[-3mm]{0mm}{8mm}
& \vvv 0^\rmT && -|\by_{n-1}|_* & 
A'_{nn} 
\end{array}
\right) \,,
\end{equation}
where we observe that $A'_{nn} = A_{nn}$.
We then calculate $\bA''$ using the similarity transformation
\begin{equation}
\bA''=\bH_{n-2}\,\bA'\,\bH_{n-2} \,.
\end{equation}
The result is 
\begin{equation}
\bA'' =
\left(
\begin{array}{ccc|c|c}
&&&0&\\
&\bB'_{n-2}&&\vdots & \vvv 0\\
&&& -|\by_{n-2}|_* &\\
\hline
\rule[-3mm]{0mm}{8mm}
0 & \cdots & -|\by_{n-2}|_* & a'_{n-1\,n-1}& -|\by_{n-1}|_* \\
\hline
\rule[-3mm]{0mm}{8mm}
& \vvv 0^\rmT && -|\by_{n-1}|_* & A'_{nn}
\end{array}
\right)\,,
\end{equation}
where
\begin{equation}
\bB'_{n-2} =\bH_{\bv_{n-2}} \; \bB_{n-2} \; \bH_{\bv_{n-2}}\,.
\end{equation}
A total of $n-2$  iterations of this process leads to a tridiagonal matrix $\bT$,
where 
\begin{equation}
\label{defT}
\bT=\bZ^{-1}\,\bA\,\bZ \,,
\end{equation}
with
\begin{subequations}
\begin{align}
\bZ= & \; \bH_{n-1} \; \bH_{n-2}\dots\bH_2 \,,
\qquad
\bZ^{-1}= \bZ^{\rm T} =
\bH_2 \; \bH_3\dots\bH_{n-1} \, .
\end{align}
\end{subequations}
In order to write a computationally efficient algorithm,
it is helpful to observe that the explicit calculation of the 
$\bZ$ matrix actually is unnecessary. 
Obviously, the only computationally nontrivial step in the 
iteration of the Householder transformations consists in the 
calculation of the matrix
\begin{align}
{\mat B}' = & {\mat H}_{\vvv v} \; {\mat B} \; {\mat H}_{\vvv v}
= \left(\mathbbm{1} -2 \, 
\frac{\vvv{v} \otimes \vvv{v}^{\rm T}}{|\vvv{v}|_*^{2}} \right) 
\; {\mat B} \; 
\left(\mathbbm{1}-2 \, 
\frac{\vvv{v} \otimes \vvv{v}^{\rm T}}{|\vvv{v}|_*^{2}} \right) 
\nonumber\\
= & {\mat B} - \vvv{v} \otimes \vvv{u}^{\rm T} -
\vvv{u} \otimes \vvv{v}^{\rm T} +
2 q \, \vvv{v} \otimes \vvv{v}^{\rm T} 
= {\mat B} - \vvv{v} \otimes \vvv{w}^{\rm T} -
\vvv{w} \otimes \vvv{v}^{\rm T} \,,
\end{align}
where we skip a few algebraic steps in the 
derivation and use the definitions
\begin{equation}
p = \frac{1}{2}|\vvv{v}|_*^{2} \,,
\quad
\vvv{u} = \frac{{\mat B} \; \vvv{v}}{p} \,,
\quad
q = \frac{\vvv{v}^{\rm T} \cdot \vvv{u}}{2p} \,,
\quad
\vvv{w} = \vvv{u}-q\vvv{v} \,.
\end{equation}
It is advantageous to calculate, for each iteration,
the vector $\vvv{v}$, then $p$, $\vvv{u}$, $q$, $\vvv{w}$ and finally
$\mat B'$.

%
%
\subsection{Diagonalization}
\label{sec2_3}

The tridiagonal matrix $\bT$ obtained in Eq.~\eqref{defT} is sparsely
populated; the only nonvanishing entries are on the diagonal, the superdiagonal
and the subdiagonal. It can be written in the form
\begin{equation}
\label{Tdef}
\bT= \left( \begin{array}{ccccc}
D_1 & E_1 &&&\\
E_1 & D_2 &E_2 &&\\
& E_2 & \ddots & \ddots &\\
&& \ddots & D_{n-1} & E_{n-1}\\
&&& E_{n-1} & D_n
\end{array}
\right).
\end{equation}
In principle, a number of methods are available for 
the diagonalization of such sparsely populated matrices.
One of these is based on QL factorization.
In its most basic version~\cite{Fr1961,Fr1962}, 
the QL factorization implements the similarity
transformations by first calculating the decomposition of a
symmetric triangular input matrix $\mat T$, as given by $\mat T = 
\bQ \, \bL$ where $\bQ$ is an orthogonal matrix 
($\bQ^{\rm T} = \bQ^{-1}$),
and $\bL$ is a lower diagonal matrix.  One then implements the
similarity transformations by simply calculating $\bT' = \bL \; \bQ = 
\bQ^{-1} \, \bT \, \bQ$.
This corresponds to an iterative
similarity transformation $\mat T = \mat Q \, \mat T' \, \mat Q^{-1}
= \mat Q \, \mat Q' \, \mat T'' \, \mat Q'^{-1} \mat Q^{-1}$ and so on.
The plain QL factorization is known to be an efficient
algorithm for wide classes of input matrices~\cite{GovL1996,Pa1998matrix}.
If the input matrix is triangular, one can
show~\cite{Wi1965,Wi1968} that the rate of convergence in 
the $K$th iteration goes
as $(\lambda_i/\lambda_{i+1})^K$, for an ordered sequence of eigenvalues
$|\lambda_1| < |\lambda_2| <  \dots < |\lambda_n|$ of an $n \times n$ input
matrix. When the matrix $\bT$ is diagonalized to machine accuracy, 
a fixed point of the similarity transformation is reached.
For complex input matrices~\cite{Wo1999,AnEtAl1999}, the common form of the 
QL decomposition calls for $\bQ$ to be unitary 
($\bQ^{-1} = \bQ^\plus$) rather than 
complex and symmetric ($\bQ^{-1} = \bQ^{\rm T}$).
Our $\bQ$ matrices have the latter property and represent 
a slight generalization of the commonly accepted version of the 
QL decomposition.

We use a so-called Wilkinson shift in order
to enhance the rate of convergence, as described in
Sec.~8.13 of Ref.~\cite{Pa1998matrix}.
The ``implicit shift'' involves a guess $\sigma$ for a specific eigenvalue of 
$\bT$, and the ensuing implementation of the 
similarity transformation $\bT \to \bT'$ is known as ``chasing the bulge''.
One performs the decomposition on a matrix shifted 
by the guess for the eigenvalue,
\begin{subequations}
\label{simtrafo}
\begin{equation}
\bT - \sigma \, \mathbbm{1}_{n \times n} = \bQ \, \bL \,,
\end{equation}
and uses the fact that 
\begin{equation}
\label{update}
\bT' = \bL \; \bQ + \sigma \, \mathbbm{1}_{n \times n}
= \bQ^{-1} \; \bT \; \bQ \,.
\end{equation}
\end{subequations}
Indeed, in a computationally efficient algorithm,
neither $\bQ$ nor $\bL$ are ever explicitly computed.
One takes advantage of the fact that the similarity 
transformation~\eqref{simtrafo} is equivalent to a series
of Jacobi~\cite{Ja1846} and Givens~\cite{Gi1958} rotations,
as described in the following.

In the first step, the implicit ``Wilkinson'' shift $\sigma$ is 
calculated~\cite{Wi1965,Wi1968}
from one of the eigenvalues of the $2 \times 2$ matrix
in the upper left corner of~\eqref{Tdef},
\begin{equation}
\left(
\begin{array}{cc}
D_{1} & E_{1}\\
E_{1} & D_2
\end{array}
\right) \,.
\end{equation}
It reads as follows,
\begin{equation}
\label{guess}
\sigma = \frac{D_2 + D_{1}}{2} 
\pm \sqrt{\left(\frac{D_2 - D_{1}}{2}\right)^2 +  E^2_{1}} \,.
\end{equation}
The $\pm$ sign is chosen such as to minimize the complex modulus 
$|\sigma - D_1|$ of the difference of $\sigma$ and the diagonal entry $D_1$.
The guess for the eigenvalue is calculated for the 
upper left corner of the input matrix,
while, in the QL decomposition,
the implicitly shifted Jacobi rotation $\bJ$ is a complex
orthogonal (but not unitary) matrix which rotates
the lower right corner of the tridiagonal input matrix
as follows, 
\begin{equation}
\bJ = \left( \begin{array}{ccccc}
1&&&& \\
 & \ddots&&& \\
  & & 1&& \\
& & & c & s \\
 & & & -s & c \end{array} \right),
\qquad \bJ^\rmT \, \bJ = \mathbbm{1}_{n\times n} \,,
\end{equation}
with manifestly complex entries 
$c^2 + s^2 = 1$ (but in general $|c|^2 + |s|^2 \neq 1$),
\begin{align}
c =& \; \frac{D_{n}-\sigma}{\sqrt{(D_{n}-\sigma)^{2}+E_{n-1}^{2}}} \,,
\qquad
s = \; \frac{E_{n-1}}{\sqrt{(D_{n}-\sigma)^{2}+E_{n-1}^{2}}} \,.
\end{align}
The transformed matrix $\bT' =\bJ^T\,\bT\,\bJ$ has the form
\begin{equation}
\bT' = \left( \begin{array}{ccccc}
\ddots&\ddots&&&\\
\ddots & D'_{n-3} & E'_{n-3} & &\\
& E'_{n-3} & D'_{n-2} & E'_{n-2} & F'\\
& & E'_{n-2} & D'_{n-1} & E'_{n-1}\\
& & F' & E'_{n-1} & D'_n
\end{array} \right)
\end{equation}
with an obvious ``bulge'' (entry $F$) at the 
elements $T'_{n-2\;n} = T'_{n\;n-2} \neq 0$.
The bulge can be ``chased upward'' using a 
generalized (complex and symmetric, but not Hermitian) Givens rotation,
\begin{equation}
\bG =
\left( 
\begin{array}{cccccc}
1&&&&& \\
& \ddots&&&& \\
& & 1&&& \\
& & & c & s& \\
& & & -s & c &\\
& & & & & 1
\end{array} 
\right),
\qquad \bG^\rmT \, \bG = \mathbbm{1}_{n\times n} \,,
\end{equation}
where again $c^2 + s^2 = 1$, and
\begin{align}
c =& \;  \frac{{E'}_{n-1}}{\sqrt{{E'}_{n-1}^{2}+{T'}_{n \, n-2}^{2}}} \,,
\qquad
s =  \frac{{T'}_{n n-2}}{\sqrt{{E'}_{n-1}^{2}+{T'}_{n \, n-2}^{2}}} \,.
\end{align}
The second transformation leads to 
$\mat T'' = \bG^\rmT \; \bT' \; \bG$ with
\begin{equation}
\bT''= \left(
\begin{array}{ccccc}
\ddots&\ddots&&&\\
\ddots&D''_{n-3}&E''_{n-3}&F''&\\
&E''_{n-3}&D''_{n-2}&E''_{n-2}&\\
&F''&E''_{n-2}&D''_{n-1}&E''_{n-1}\\
&&&E''_{n-1}&D''_n
\end{array}
\right).
\end{equation}
Upon a Givens rotation, one updates the entries 
on the diagonal and sub-(super-)diagonal 
of the tridiagonal matrix $\bT \to \bT' \to \bT''$.
The additional element of the ``bulge'' can be stored as 
a single variable.
After $n-2$ (Givens) $\bG_j$ rotations with $j=n-2, \dots 1$,
starting from $\bG_{n-2} \equiv \bG$ and continuing to $\bG_1$, 
the bulge has disappeared, and $\bT$ again assumes a tridiagonal form.
The orthogonal transformation $\bQ$ from Eq.~\eqref{update} is 
identified as
\begin{equation}
\bQ = \bJ \; \bG_{n-2} \; \bG_{n-3} \cdots \bG_1 \,.
\end{equation}
In general, the convergence toward the 
eigenvalues in the $K$th iteration is improved~\cite{Wi1965,Wi1968}
to $[(\lambda_i -\sigma)/(\lambda_{i+1}-\sigma)]^K$, 
again for an ordered sequence of eigenvalues 
$|\lambda_1| < |\lambda_2| <  \dots < |\lambda_n|$.
The similarity transformations are iterated
until the off-diagonal element $E_1$ is zeroed
to machine accuracy.
One then repeats the process for the lower right 
$(n-1) \times (n-1)$ submatrix of $\bT$,
then, for the 
$(n-2) \times (n-2)$ submatrix of $\bT$,
each time zeroing the first off-diagonal element,
until $\bT$ is diagonalized to machine accuracy.

\begin{figure*}[th!]
\begin{center}
\begin{minipage}{0.99\textwidth}
\begin{center}
\includegraphics[width=1.0\linewidth]{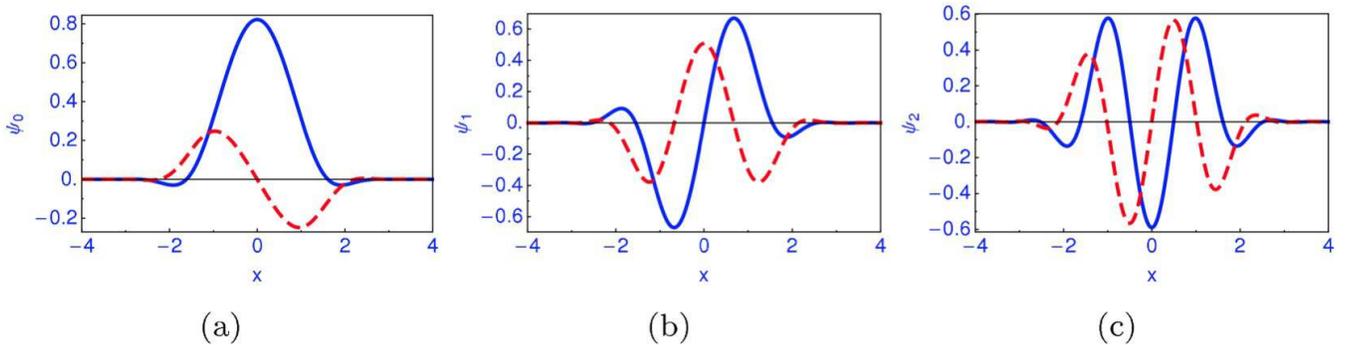}
\end{center}
\caption{\label{fig1}{(Color online.)
Figure~(a) displays the ground-state wave function of the 
imaginary cubic Hamiltonian~\eqref{H3} for $G = 1.0$.
This wave function is manifestly complex.
The real part $\mathrm{Re} \, \psi_0(x)$ 
(even under parity)
is plotted using solid lines, 
and the dashed curve corresponds to the 
parity-odd imaginary part
$\mathrm{Im} \, \psi_0(x)$.
For the first excited state (still, $G = 1.0$),
the real part is odd, while the imaginary part is even under parity
[see Fig.~(b)].  The second excited state [Fig.~(c)] has an
even real part, while its imaginary part is odd.
The global complex phase of the wave function 
is chosen so that the real part $\mathrm{Re} \, \psi_{n=0,1,2}(x)$
of the wave functions has the same qualitative behavior
as the eigenstate wave function of the quartic oscillator 
displayed in Fig.~\ref{fig2}(a).}}
\end{minipage}
\end{center}
\end{figure*}

%
%
\subsection{Numerical reference data}
\label{sec2_4}

The algorithmic procedure described above leads to a matrix diagonalization
algorithm for complex symmetric matrices, which can find
both the eigenvalues and eigenvectors of the original input matrix~$\bA$.
We have checked numerical results obtained for wide classes
of \PT-symmetric anharmonic oscillators
against numerous published data.
An example of an interesting 
alternative procedure for the calculation of the eigenenergies
is given by the moment method~\cite{HaBe1985,HaBeMo1988,HaBeSiMo1988},
which relies on a Fourier transformation of the
Schr\"{o}dinger equation and is based on a recursive calculation of the
moments which define the series expansion of the
wave function in momentum space.
Applied to the positive-definite measure $S(x) = |\psi(x)|^2$,
the method has been shown to generate numerical approximations for the
discrete states of the $-{\rm i}\,x^3$ non-Hermitian potential~\cite{Ha2001letter},
as well as a potential proportional to $\ii\,x^3+\ii\,\alpha\,x$,
which induces \PT\,-symmetry breaking, manifestly complex 
eigenvalues~\cite{HaKhWaTy2001,Ha2001long}.
The wave functions of eigenstates have also been studied,
including Stokes and Anti-Stokes lines, for both the imaginary cubic
oscillator~\cite{HaWa2001} as well as generalized
$(\ii\,x)^N$-potentials~\cite{YaHa2001}.
Furthermore, we have used the algorithm for the 
calculation of resonance and anti-resonance energies 
of the ``real'' cubic perturbation (potential proportional to $x^3$)
and other odd anharmonic oscillators.
We note that the eigenvalues of the \PT-symmetric imaginary cubic perturbation and the 
Hermitian, but not essentially self-adjoint real cubic oscillator
are related by a dispersion relation~\cite{JeSuZJ2009prl,BeDu1999}.

For reference, let us consider the two Hamiltonians
\begin{subequations}
\label{hams}
\begin{align}
\label{hama}
h_3 =& \; -\frac12 \, \ee^{-2 \ii \theta} \, \partial_x^2 + 
\frac12 \, \ee^{2 \ii \theta} \, x^2 + 
\ee^{3 \ii \theta} \, x^3 \,, 
\qquad
0 < \theta < \frac{\pi}{5} \,,
\\[2ex]
\label{hamb}
H_3 =& \; -\frac12 \, \partial_x^2 + \frac12 \, x^2 + \ii \, x^3 \,.
\end{align}
\end{subequations}
The first of these involves a complex scaling transformation,
which gives rise to manifestly complex resonance energy eigenvalues.
The complex scaling transformation is ``dual'' to the resummation
of the perturbation series to the complex resonance energies, 
which has been discussed in Refs.~\cite{FrGrSi1985,Je2001pra,Ca2003}.
Using a multi-precision arithmetic implementation~\cite{Ba1994tech,Ba1995} of the 
algorithm described in Sec.~\ref{sec2}, we easily obtain the first two
resonance energy eigenvalue of $h_3$ as follows,
\begin{subequations}
\begin{align}
\epsilon_0 = 0.&61288\,84333\,07754\,62425\,88175\,01988\,%
65141\,37333\,39788\,30718\,29420\,66181
\nonumber\\
  -          0.&40859\,26669\,32267\,28315\,94988\,68767\,
                16051\,62709\,74834\,43840\,39990\,97532 \, \ii \,,
\\[2ex]
\epsilon_1 = 2.&18041\,38375\,36348\,77123\,01619\,63541\,%
                74113\,12471\,72136\,83505\,89744\,59041
\nonumber\\
- 1.&52620\,76556\,93032\,51000\,68539\,46967\,%
     49562\,44459\,06099\,84880\,44103\,55220 \, \ii \,.
\end{align}
\end{subequations}
All of the given decimals are significant; the 60-figure precision 
is obtained in a basis of roughly 1000 harmonic oscillator eigenstates
and can be enhanced if desired.
The ground-state-energy of the Hamiltonian $H_3$ and its 
first-excited-state energy reads as follows,
\begin{subequations}
\begin{align}
E_0 = 0.&79734\,26075\,08906\,18903\,90809\,60791\,%
01316\,30972\,44534\,48033\,11575\,78578 \,,
\\[2ex]
E_1 = 2.&77352\,49851\,95379\,71540\,58170\,00015\,%
53014\,23108\,48902\,82968\,52057\,22959 \,.
\end{align}
\end{subequations}
We again reemphasize that the precision of these results can 
be easily enhanced, as it was necessary to test some of the 
conjectured generalized quantization conditions for anharmonic 
oscillators presented in 
Refs.~\cite{JeSuLuZJ2008,JeSuZJ2009prl,JeSuZJ2010,ZJJe2010jpa}.
For typical 
double-precision (16 decimals) and 
quadruple-precision (32~decimals) calculations, we observe that the 
timings for matrix diagonalization using our algorithm 
are comparable to those using the routine {\tt ZGEEVX} 
built into the {\tt LAPACK} library~\cite{AnEtAl1999}.
Using a dedicated, concise implementation of the algorithm
of the algorithm discussed in Sec.~\ref{sec2_2} and Sec.~\ref{sec2_3}, 
we were even able to obtain timings which exceed the speed of {\tt LAPACK}'s 
by up to 50\% for typical applications (matrices of rank $500 \times 500$);
however, the speed-up may be compiler-specific (we were using {\tt gfortran}
version 4.7.2, see Ref.~\cite{gfortran}).

%
%
\section{Pseudo--Hermitian Quantum Mechanics}
\label{sec3}

%
%
\subsection{Orientation}
\label{sec3_1}

Let us try to explore the physical motivation for the 
construction of the matrix diagonalization algorithm
presented in Sec.~\ref{sec2}, on the basis of pseudo-Hermitian
(\PT-symmetric) quantum mechanics.
A \PT-symmetric Hamilton operator $H$ fulfills the relation
\begin{equation}
\label{relationH3}
H = \calP \; \calT \; H \; \calT \; \calP
= \calP \; H^\plus \; \calP \,,
\end{equation}
where $H^\plus$ is obtained~\cite{BeBrReRe2004} from $H$ by the replacement
$\ii \to -\ii$, which is the same as the Hermitian
adjoint if all other terms in the Hamiltonian are explicitly
real (rather than complex).
The parity and time reversal are denoted as $\calP$ and $\calT$, respectively.
If a Hamiltonian $H$ fulfills a relation of the type
$H = \eta^{-1} \, H^\plus \, \eta$,
then $H$ is said to be pseudo-Hermitian~\cite{Pa1943}.
\PT-symmetry can thus be interpreted as a special case of
pseudo-Hermiticity ($\eta = \calP$), even if there is a certain 
``clash'' with the original definition from Ref.~\cite{Pa1943},
where it was assumed that $\eta$ is a positive-definite operator.
By contrast, $\calP$ may have the negative eigenvalue $-1$.

For reference, we continue our analysis with the 
well-known imaginary cubic 
perturbation~\cite{JeSuLuZJ2008,JeSuZJ2009prl,JeSuZJ2010,ZJJe2010jpa,BeBo1998,BeBoMe1999},
which we had already employed in Sec.~\ref{sec2_4}.
It is described by the Hamiltonian 
\begin{equation}
\label{H3}
H_3 = - \frac12 \, \partial_x^2 + \frac12 \, x^2 + \ii \, G \, x^3 \,,
\qquad G > 0 \,,
\end{equation}
which for $G=1$ reduces to Eq.~\eqref{hamb}.
The eigenfunctions of $H_3$ are manifestly complex,
in contrast to those of the quartic anharmonic oscillator,
\begin{equation}
\label{H4}
H_4 = - \frac12 \, \partial_x^2 + \frac12 \, x^2 + g \, x^4\,,
\qquad g > 0 \,,
\end{equation}
where the eigenstate wave functions can be chosen as purely real.
For a \PT-symmetric system, the scalar product 
\begin{equation}
\label{star_scalar_prod}
\langle\psi(t) | \phi(t) \rangle_{\mathcal{P}\mathcal{T}} \equiv
\int \dd x \, \psi^*(x,t) \, \calP \phi(x,t) 
\end{equation}
is conserved under time evolution if both $\psi$ and $\phi$
fulfill the time-dependent Schr\"{o}dinger equation 
$\ii \partial_t \psi(t) = H_3 \psi(t)$, and
$\ii \partial_t \phi(t) = H_3 \phi(t)$.
However, the integrand in Eq.~\eqref{star_scalar_prod} is manifestly 
``nonlocal'' because $ \psi^*(x,t) \calP \phi(x,t)  =
\psi^*(x,t) \phi(t,-x)$; it depends on function 
evaluations at $x$ and $-x$.
This is in contrast to the ordinary scalar product
\begin{equation}
\label{ordinary_scalar_prod}
\langle\psi(t) | \phi(t) \rangle \equiv
\int \dd x \, \psi^*(x,t) \, \phi(x,t) \,,
\end{equation}
where the first argument is complex conjugated,
and the (generalized) indefinite inner product
\begin{equation}
\label{complex_inner_prod}
\langle\psi(t) | \phi(t) \rangle_* \equiv
\int \dd x \, \psi(x,t) \, \phi(x,t) \,,
\end{equation}
where none of the arguments are complex conjugated.
The Hamiltonian $H_3$ given in Eq.~\eqref{H3} involves a 
manifestly complex potential, which we denote as $W(x)$,
\begin{subequations}
\begin{align}
\label{Vcomplex}
W(x) =& \; \frac12 \, x^2 + \ii \, G \, x^3 = 
V(x)\, \ee^{\ii \, \arg(V(x))} \,,
\\[2ex]
V(x) =& \; |W(x)| = \sqrt{ \tfrac14 \, x^4 + G^2 \, x^6} \,.
\end{align}
\end{subequations}
The modulus $V(x) = |W(x)|$ 
tends to infinity as $x \to \pm \infty$.
For purely real  potentials like the ``confining'' quartic 
oscillator given in Eq.~\eqref{H4},
intuition suggests that the ``bulk'' of the 
probability density of the eigenstate wave function 
should be concentrated in the ``classically allowed'' region,
i.e., in the region where the eigenenergy $E$ is greater than
the potential, i.e., $E > V(x)$ [where $V(x) \in \mathbbm{R}$]. 
For a manifestly complex
potential, the condition $E > W(x)$ 
[with $W(x) \in \mathbbm{C}$] cannot be applied
because the complex numbers are not ordered. 
In the following, we consider a few example
calculations of energy eigenvalues of the 
imaginary cubic perturbation~\eqref{H3},
which illustrate these observations.
All of these have been accomplished using the algorithm 
presented in Sec.~\ref{sec2}.

\begin{figure*}[t!]
\begin{center}
\begin{minipage}{0.99\textwidth}
\begin{center}
\includegraphics[width=1.0\linewidth]{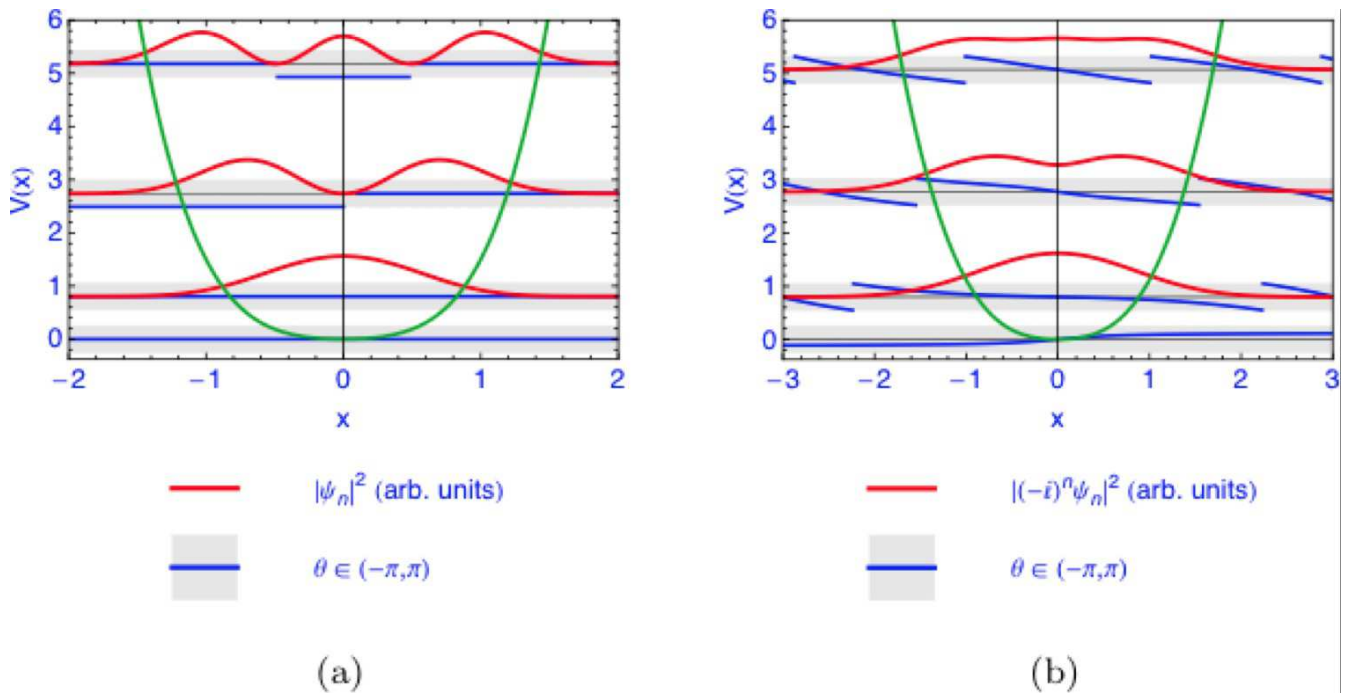}
\end{center}
\caption{\label{fig2}(Color online.)
In Fig.~(a), we plot the probability density $\rho=|\psi(x)|^2$ of the
quartic oscillator's ground state and first two excited states,
in a potential $V(x) = \tfrac12 x^2 + g\,x^4$ with $g=1.0$.
Although the wave functions of the 
quartic potential are purely real,
we use a modulus-phase plot for the real 
wave functions in Fig.~(a). A sign change then corresponds to a
jump in the complex phase from zero
(for a positive real number) to $-\pi$
(for a positive negative number).
In Fig.~(b), we give a modulus-phase plot of the eigenstate
wave functions of the imaginary cubic perturbation,
where the complex phase of the wave function is displayed in the shaded
region. The complex phase $\theta=\theta(x)$ in the decomposition
$(-\ii)^n \, \psi_n(x) = |\psi_n(x)|\, \exp[\ii \, \theta(x)]$ 
covers the interval $[-\pi, \pi)$.
Here, the $\psi_n$ are the wave functions
of Fig.~\ref{fig1}, multiplied by a phase factor $(-\ii)^n$
[see also Eq.~\eqref{PTunity}].}
\end{minipage}
\end{center}
\end{figure*}

\begin{figure}[t!]
\begin{center}
\begin{minipage}{0.99\linewidth}
\begin{center}
\includegraphics[width=0.45\linewidth]{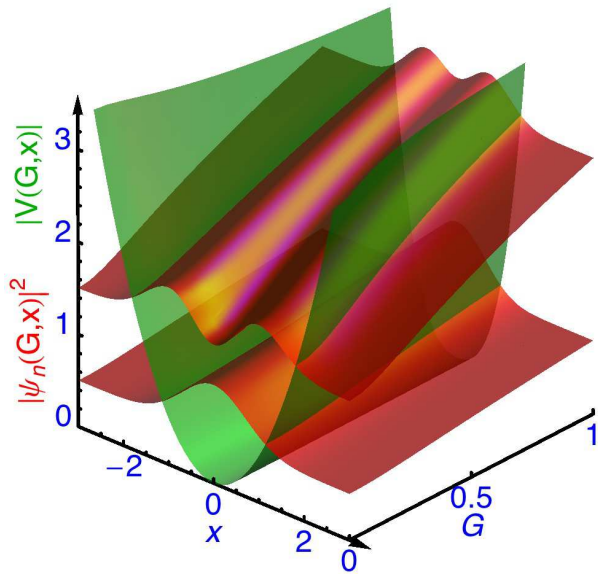}
\end{center}
\caption{\label{fig3}(Color online.) Illustration of the confinement
mechanism for the imaginary cubic potential described by the 
Hamiltonian~\eqref{H3}, for the ground and the 
first excited state. The bulk of the 
modulus square of the wave function is centered in the ``allowed''
region where the (real rather than complex) energy
$E > V(x) = V(G, x) = |W(G, x)|$. The potential is plotted 
in green (the ``trough''-like structure), 
whereas the moduli of the wave functions 
are plotted in red (the ``wave-like structures'').
The squares of the moduli of the wave functions either
have a single maximum (ground state), or two maxima (first excited state).
The ground state wave function has a 
modulus square $|\psi_0(x)|^2 = |\psi_0(G,x)|^2$ as a function
of $G$ and $x$. As $G$ increases, the bound-state energy 
(which is equal to the base line of the wave function curve
at any given value of $G$) increases,
and the modulus of the potential forms a more narrow trough
to which the ground-state wave function is confined.
The same is true for the first excited state.
The central minimum of the modulus square of the 
first-excited state wave function is clearly visible.}
\end{minipage}
\end{center}
\end{figure}

%
%
\subsection{Example calculations}
\label{sec3_2}

We can formally split the Hamiltonian $H_3$
into a ``real part'' and a ``imaginary part'' as follows,
\begin{align}
\label{decomp}
\mathrm{Re} \, H_3 = & \; - \half \, \partial_x^2 + \half x^2  \,,
\qquad
\mathrm{Im} \, H_3 = \ii \, G \, x^3 \,.
\end{align}
Likewise, we can also split the eigenstate wave function $\psi_n(x)$ 
into real and imaginary parts, 
\begin{equation}
\psi_n(x) = \mathrm{Re} \, \psi_n(x) + 
\ii \, \mathrm{Im} \, \psi_n(x) \,.
\end{equation}
Based on the decomposition~\eqref{decomp}, one can show
that if $\mathrm{Re} \, \psi_n(x)$ is even under parity
and $\psi_n(x) $ is an eigenstate of $H_3$ with 
real eigenvalue of $\epsilon_n$,
then $\mathrm{Im} \, \psi_n(x)$ has to be parity-odd,
and vice versa.
Also, if $\mathrm{Re} \, \psi_n(x)$ is odd under parity,
then $\mathrm{Im} \, \psi_n(x)$ has to be even under parity.
Because the parity operator $\calP$ does not
commute with the Hamiltonian
$H_3$, the eigenstates of $H_3$ are not eigenstates of parity.
Furthermore, because the potential is manifestly
complex, so are the wave functions.
Yet, numerical evidence drawn from Fig.~\ref{fig1}
suggests that if the global phase of the 
wave function is appropriately chosen,
both real as well as imaginary
parts of the eigenstates are eigenstates of parity,
individually. These eigenstates are naturally obtained if 
one diagonalizes an approximation to the cubic Hamiltonian
obtained by projection onto a suitably large
basis set of harmonic oscillator eigenstates.

The antilinear $\mathcal{P}\mathcal{T}$ 
operator commutes with the Hamiltonian, and the 
eigenfunctions of $H_3$ are also
$\mathcal{P}\mathcal{T}$ eigenstates~\cite{BeBrReRe2004}.
The precise eigenvalue of \PT{} may, however, depend on the 
phase assigned to $\psi_n(x)$ because $\calT$ is an antilinear 
operator. Let us first investigate the phase conventions 
used in Fig.~\ref{fig1}, where the real and imaginary 
parts of the wave function are, alternatingly, even and odd under 
parity as we proceed to higher excited states.
The appropriate eigenvalues are thus
\begin{equation}
\mathcal{P}\mathcal{T} \psi_n(x) = 
\psi^*_n(-x) = (-1)^n \, \psi(x) \,.
\end{equation}
However, the eigenvalue of the 
wave functions $\Psi_n(x) = (-\ii)^n \psi_n(x)$
with respect to the $\mathcal{P}\mathcal{T}$ operator
is unity,
\begin{equation}
\label{PTunity}
\mathcal{P}\mathcal{T} [(-\ii)^n \, \psi_n(x)] = 
\ii^n \, \psi^*_n(x) = [ (-\ii)^n \, \psi(x)] \,.
\end{equation}
For two eigenstates of the \PT-symmetric system,
the conserved \PT-symmetric scalar product 
simply is the indefinite inner product,
as defined in Eq.~(2.4.2) of Ref.~\cite{Mo1998},
\begin{align}
\label{IDscal}
\langle\Psi_n | \Psi_m \rangle_{\mathcal{P}\mathcal{T}} =& \;
\int \dd x \, \Psi_n^*(x) \calP \Psi_m(x) 
= \int \dd x \, \left(\calP\calT \Psi_n\right)(x) \, \Psi_m(x) 
\nonumber\\
=& \; \int \dd x \, \Psi_n(x) \, \Psi_m(x) = 
\langle\Psi_n | \Psi_m \rangle_* \,.
\end{align}
The integrand in Eq.~\eqref{IDscal} is 
``local'' in the sense that it depends only on 
wave functions at $x$, not $-x$.
The non-local character of the 
integrand in Eq.~\eqref{star_scalar_prod}
has otherwise been called into question and has given 
rise to rather sophisticated attempts at finding 
an alternative, appropriate interpretation~\cite{AbJaPa2013}.
The natural normalization condition for eigenstates
of complex, symmetric matrices 
(including infinite-dimensional matrices) is given by~\cite{Mo1998}
\begin{equation}
\label{equality}
\langle\psi_n | \psi_m \rangle_*  = \delta_{nm} \,,
\qquad
\langle\Psi_n | \Psi_m \rangle_*  = (-1)^n \, \delta_{nm} \,,
\end{equation}
and involves the indefinite inner product defined in 
Eq.~\eqref{complex_inner_prod}.
The equivalence shown in Eq.~\eqref{IDscal} and the 
orthogonality properties~\eqref{equality} provide the 
main motivation for the construction of the 
generalized Householder transformation~\eqref{gen_house};
indeed, the generalized inner product~\eqref{IDscal}
reduces to~\eqref{gen_inner} for finite-dimensional vector spaces.
``Half'' of the $\Psi_n$ eigenstates acquire a negative 
\PT-symmetric norm, as described by the prefactor $(-1)^n$.
For two time-dependent states of the \PT-symmetric system, given as
\begin{equation}
\label{chirho}
\chi(t) = \sum_n a_n(t) \; | \Psi_n \rangle \,,
\quad
\rho(t) = \sum_m b_m(t) \; | \Psi_m \rangle \,,
\end{equation}
with $a_n = (-1)^n \, \langle \Psi_n | \chi \rangle$,
and $b_n = (-1)^n \, \langle \Psi_n | \rho \rangle$,
the \PT-symmetric scalar product is calculated as follows,
\begin{align}
\label{mixed}
\langle \chi(t) | \rho(t) \rangle_{\mathcal{P}\mathcal{T}} =& \;
\sum_{nm} a^*_n(t) \, b_m(t) \,
\int \dd x \, \Psi_n^*(x) \calP \Psi_m(x)
= \sum_{nm} a^*_n(t) \, b_m(t) \,
\int \dd x \, \Psi_n(x) \Psi_m(x) 
\nonumber\\
=& \; \sum_{n} (-1)^n \, a^*_n(t) \, b_n(t) 
= \langle \vec a(t) | \vec b(t) \rangle_{\rm even} 
- \langle \vec a(t) | \vec b(t) \rangle_{\rm odd} \,.
\end{align}
with an obvious identification of 
$\langle \vec a(t) | \vec b(t) \rangle_{\rm even}$ and 
$\langle \vec a(t) | \vec b(t) \rangle_{\rm odd}$.
Note that one cannot suppress the factor $(-1)^n$
in the second-to-last line of Eq.~\eqref{mixed}
by a change in the global phase factor of the wave functions.
The factor  either occurs because of the \PT-symmetric eigenvalue 
of the $\psi_n$, or because of the 
alternating sign of the norm of the $\Psi_n$.
The \PT-symmetric time evolution is separately 
unitary in the space of the coefficients 
$a_n(t)$, $b_n(t)$ with (i)~even $n$ and 
(ii)~odd $n$, as denoted by an appropriate subscript in 
Eq.~\eqref{mixed}. In some sense, 
the \PT-symmetric time evolution leads to a natural 
``splitting'' of the Hilbert space into two 
subspaces, those of the function with negative 
\PT-symmetric norm and those with positive norm,
according to the second equality in Eq.~\eqref{equality}.
The same pattern has recently been observed in a 
field-theoretical context~\cite{JeWu2012epjc,JeWu2013isrn}: Half of the 
states of the generalized Dirac equation with a 
pseudo-scalar mass term acquire a negative Fock-space norm.

From ordinary, Hermitian, quantum mechanics,
it is known that the $L^2(\mathbbm{R})$ eigenfunctions 
of a Hermitian operator are in some sense 
confined to spatial regions where the 
eigenenergy $E_n$ is larger than the local value of the 
potential, $E > V(x)$.
An intuitive understanding can be obtained if we interpret
the potential in terms of a modulus and a phase,
according to Eq.~\eqref{Vcomplex}.
We use the rationale of accompanying a 
plot of the modulus of a function by a grey band 
to convey complex phase information in Figs.~\ref{fig2}
and~\ref{fig3}, where the complex eigenstate wave functions 
shown previously in Fig.~\ref{fig1}
are plotted in terms of the decomposition
\begin{equation}
\psi_n(x) = | \psi_n(x) | \, \ee^{\ii \, \arg(\psi_n(x))} \,.
\end{equation}
The curves in Figs.~\ref{fig2} and~\ref{fig3} show that indeed,
the wave functions of the \PT-symmetric 
oscillator are concentrated [in the sense of a 
large absolute value of the integrand in 
Eq.~\eqref{IDscal}] to a region where $E_n > V(x) = |W(x)|$,
where $W(x)$ is the complex-valued \PT-symmetric 
potential~\eqref{Vcomplex}.
In Fig.~\ref{fig3}, we illustrate that the ``confinement''
mechanism holds for all values in the range $0 < G < 1$.
The interlacing of zeros in complex Sturm--Liouville
problems for \PT-eigenfunctions,
has been discussed in Ref.~\cite{BeBoSa2000}.
In agreement with the conclusions of Ref.~\cite{BeBoSa2000},
we find that both the real as well as imaginary 
parts of the complex eigenfunctions have an 
infinite number of zeros, individually,
when the argument $x$ of the wave function covers 
the real numbers.
Furthermore, this statement even holds for an infinitesimally small,
but nonvanishing coupling $G$ in the imaginary cubic perturbation
$\ii \, G  \, x^3$.

%
%
\section{Conclusions}
\label{sec4}

In Sec.~\ref{sec2}, we have presented an efficient algorithm for the
calculation of eigenvalues of complex symmetric (not Hermitian) matrices. The
algorithm is scalable in terms of the desired numerical accuracy and relies on
generalized Householder reflections which ``depopulate'' the input matrix by
projecting the entries onto the sub- and super-diagonals, using the generalized
inner product which avoids the complex conjugation of the first argument.  We
find that a subsequent diagonalization of the tridiagonal matrix obtained from
the Householder transformations, using generalized Jacobi and Givens rotation
matrices (which are again complex and symmetric but not unitary) leads to an
efficient eigenvalue solver.  Numerical reference data are provided in
Sec.~\ref{sec2_4}, and we reemphasize that many of the previously reported
numerical tests of generalized quantization conditions for anharmonic
oscillators~\cite{JeSuLuZJ2008,JeSuZJ2009prl,JeSuZJ2010,ZJJe2010jpa} rely on
the numerical methods described in this paper.
The indefinite inner product can indeed be useful in numerical algorithms;
in Ref.~\cite{ArHo2004}, the indefinite inner product had been used previously
in a reformulation of the Rayleigh quotient,
within an adaptation of the Jacobi--Davidson method for complex
symmetric matrices (which has nothing to do with the 
Jacobi rotations used in our algorithm). We might add that
in contrast to the Jacobi--Davidson method,
our algorithm does not require a Gram--Schmidt orthogonalization step and
is based on a generalized inner product which draws its inspiration
from physics.

We would like to illustrate and comment on the algorithm by pointing out a few
possible modifications and intricacies of the methods used.  The above version
of the algorithm described in Sec.~\ref{sec2} is based on the QL rather than QR
decomposition.  In the (shifted) iterated QL decomposition, one starts the
calculation of the eigenvalue guess from the upper left corner of the input
matrix but calculates the Givens and Jacobi rotations from the lower right,
i.e., one ``chases the bulge upward''.  In that case, the ``uppermost''
eigenvalue of the input matrix converges first; the guess $\sigma$ approximates
the true eigenvalue to machine accuracy.  If the complex symmetric Hamiltonian
is obtained from a basis set of ``trial'' quantum states the first of which
approximates the state of lowest energy of the perturbed system, then the QL
decomposition as opposed to the QR decomposition ensures that the
``ground-state energy converges first''.  Still, it is an instructive exercise
to modify the algorithm so that the ``eigenvalue guess'' $\sigma$ is first
calculated for the ``lower rightmost'' eigenvalue.  One then calculates the
Jacobi and Givens rotations which ``chase the bulge'' from the upper left to
the lower right of the tridiagonal matrix.  In that case, the ``lower
rightmost'' eigenvalue converges first.  This may be useful in particular cases
where the ``highest'' eigenvalue is of particular interest.

One has a few options for controlling the convergence of the algorithm: For
example, in many applications, the element $D_i$ may be declared to have
``converged'' if $D_i + E_i$ equals $D_i$ to machine accuracy, but this
criterion may be too restrictive in some cases, especially, when
multi-precision arithmetic is being used. In that case, it should be replaced
by a criterion which states that $| D_i/E_i|$ is less than a specific,
predefined accuracy, say $10^{-64}$ (for so-called ``octuple precision''
arithmetic, with 64~decimals).

Our QL factorization involves manifestly complex symmetric $\bQ$ matrices.
Typically, routines built into modern computer algebra system use a unitary
matrix $\bQ$ for such decompositions. These routines use manifestly different
matrices than those employed in the approach described above and therefore cannot
be used, say, in a meaningful comparison to an implementation of the above
algorithm.

In the iterated, shifted QL decomposition of the tridiagonal matrix $\bT$, the
most common pitfall consists in a ``premature zero'', i.e., in an entry on the
sub- or super-diagonal $E_j$ which becomes zero to machine accuracy before the
``target element'' $E_{i<j}$ for which the current guess $\sigma$ is calculated
has converged to the desired accuracy.  In that case, the tridiagonal matrix
naturally divides into two matrices (two ``irreducible representations'') which
have to be considered separately.  Typically, this phenomenon occurs when the
entries in the original input matrix $\bA$ have a somewhat irregular pattern
(e.g., random matrices).  The necessity to partition the tridiagonal matrix
upon the occurrence of premature zeroes is described rather scarcely in the
literature; some lecture notes on the matter can be found in Sec. 11.4 of
~\cite{Fa2006} and near the end of Sec. 3.6.2 of Ref.~\cite{Ar2012}. The
division into two matrices is called ``partitioning'' in Sec. 4.7 of
~\cite{He1991}.  For matrices obtained from regularly distributed, trial basis
states, which typically occur in theoretical physics, we have not observed this
phenomenon.

As a final remark, we would like to mention that a plain iterated QL or QR
decomposition leads to a rather efficient, but not optimized, convergence of
the eigenvalues problem, especially for regular input entries in the matrix
$\bA$.  The QL and QR decompositions of the input matrix $\bA$ can be
calculated using generalized Householder reflections: For QL, one starts from
the rightmost column vector of $\bA$ and projects it onto its last element; for
QR, on starts from the leftmost column vector of $\bA$ and projects it onto its
first element; the subsequent Householder reflections are constructed from the
``deflated'' $(n-1) \times (n-1)$ submatrices, in either
direction~\cite{GovL1996,Pa1998matrix}.  Skipping the tridiagonalization step,
one can thus construct generalized QL and QR factorization-based matrix
diagonalization routines where $\bQ$ is manifestly complex and symmetric, but
not unitary.  Otherwise, the implicit shift~\eqref{simtrafo} leads to improved
convergence toward to numerical approximations of the eigenvalues.

The usefulness of the generalized Householder reflections used in the algorithm
has a connection to the underlying physics.  Indeed, we find that the most
natural interpretation of the conserved scalar product in the \PT-symmetric
time evolution is in terms of the generalized indefinite inner product defined for
finite-dimensional vectors in Eq.~\eqref{gen_inner} and for Hilbert space
vectors in Eq.~\eqref{complex_inner_prod}.  This inner product is linear in
both arguments and avoids complex conjugation.  Attempts at finding an
alternative, appropriate interpretation~\cite{AbJaPa2013} seem too complicated
to take precedence over the immediate identification of the \PT-symmetric
scalar product of eigenvectors in terms of the indefinite inner product. The
integrand immediately becomes ``local'' and one avoids integrations over
eigenfunctions evaluated at the (possibly very distant) points $x$ and $-x$.
We reemphasize that the indefinite inner product is crucial in the analysis of
resonance eigenvectors~\cite{Mo1998} for the ``real'' cubic
perturbation, and that the eigenvalues of the ``real'' and ``imaginary'' cubic
perturbation are related by a dispersion
relation~\cite{BeDu1999,JeSuZJ2009prl}.  So, in some sense, the importance of
the indefinite inner product for the analysis of the imaginary cubic Hamiltonian
had to be expected, and the emergence of an efficient matrix diagonalization
algorithm for such matrices is only natural (see Sec.~\ref{sec3}).

It has been stressed in the literature that the scalar product $\langle \psi(t)
| \phi(t) \rangle_{\calP\calT}$ as defined in Eq.~\eqref{star_scalar_prod} is
not positive definite. This has been used as an argument against the viability
of \PT-symmetric Hamiltonians for the description of natural phenomena.
However, one may counter argue that the same problem persists with regard to
the relativistic Klein-Gordon equation where the time-like component of the
conserved Noether current can become negative~(see Chap.~2 of
Ref.~\cite{ItZu1980}).  The Klein-Gordon equation is assumed to describe a
charged scalar field like the charged component of the Higgs (doublet)
field~\cite{PeSc1995} (the latter is usually assumed to vanish under a gauge
transformation, and the remaining neutral component of the Higgs doublet is
expanded about its vacuum expectation value). Strictly speaking, one has to
reinterpret the timelike component of the conserved Noether current as a charge
density, not a probability density.  Analogously, the 
 timelike component of the conserved Noether current 
of the pseudo--Hermitian, generalized Dirac Hamiltonian (with a pseudo-scalar
mass term) may naturally be interpreted as a 
non-positive definite ``weak-interaction density'' (see Ref.~\cite{JeWu2013isrn}).
Equation~\eqref{mixed} suggests that the Hilbert space, under the \PT-symmetric
time evolution, is split into two ``halves'', one of which entails negative
\PT-symmetric norm (analogous to the ``right-handed neutrinos''), and the other
has positive \PT-symmetric norm (analogous to the ``left-handed neutrinos''
within the model proposed in Ref.~\cite{JeWu2013isrn}).  We recall that
``half'' of the $\Psi_n$ eigenstates acquire a negative \PT-symmetric norm
under a very natural choice of the global complex phase [see
Eq.~\eqref{chirho}]. The \PT-symmetric norm, in turn, can be 
formulated in terms of the generalized indefinite inner product,
on which this article is based.

%
%
\section*{Acknowledgments}

Support by the National Science Foundation (Grant PHY-1068547) 
and by the National Institute of Standards and Technology 
(Precision Measurement Grant) is gratefully acknowledged.
The authors also acknowledge helpful conversations with 
Istvan Nandori and Andras Kruppa.

\end{document}